\documentclass[12pt]{article}
\usepackage{graphicx}
\usepackage{cite}

\newcommand{\be}{\begin{equation}}
\newcommand{\ee}{\end{equation}}
\newcommand{\bea}{\begin{eqnarray}}
\newcommand{\eea}{\end{eqnarray}}
\newcommand{\smallw}{{\scriptscriptstyle W}} %
\newcommand{\smallr}{{\scriptscriptstyle R}} %
\newcommand{\smalll}{{\scriptscriptstyle L}} %
\newcommand{\smallx}{{\scriptscriptstyle X}} %
\newcommand{\smallmax}{{\rm\scriptstyle max}} %
\newcommand{\smallmin}{{\rm\scriptstyle min}} %
\newcommand{\gl}{g_\smalll}
\newcommand{\gr}{g_\smallr}
\newcommand{\fl}{f_\smalll}
\newcommand{\fr}{f_\smallr}
\newcommand{\flr}{f_{\smalll \smallr}}
\newcommand{\fx}{f_\smallx}

\newcommand{\flov}{\bar{f}_\smalll}
\newcommand{\frov}{\bar{f}_\smallr}
\newcommand{\flrov}{\bar{f}_{\smalll \smallr}}
\newcommand{\fxov}{\bar{f}_\smallx}

\newcommand{\mw}{M_\smallw}

\def \sss  {\scriptscriptstyle}

\def \gev  {\mbox{ GeV}}
\def \mev  {\mbox{ MeV}}
\def \dilog  {\mbox{Li}_2}

\def \psl  {p \kern-.45em{/}}
\def \qsl  {q \kern-.45em{/}}
\def \lsim {\raisebox{-.7ex}{$\stackrel{\textstyle <}{\sim}\,$}}

\def \Be7  {$\!\!\!\!\phantom{A}^7$Be~}
\def \B8   {$\!\!\!\!\phantom{A}^8$B~}
\def \bm   {\boldmath}
\def \eom  {E_\omega}
\def \zom  {z_\omega}
\def \notin {\in \!\!\!\!\!/}

\begin{document}              

\begin{titlepage}
\begin{flushright}
        \small
        BUTP 2001/02\\
        February 2001
\end{flushright}

\renewcommand{\thefootnote}{\fnsymbol{footnote}}

\begin{center}
\vspace{2cm}
{\LARGE \bf QED Corrections to the Scattering \\
of Solar Neutrinos and Electrons\footnote{Presented at the
  Symposium in Honor of Professor Alberto Sirlin's $70^{th}$ Birthday:
        {\em 50 Years of Precision Electroweak Physics}, 
        New York University, October 27--28, 2000.}}

\vspace{1cm}
{\large\bf    M.~Passera}
\setcounter{footnote}{0}
\vspace{.5cm}

{\it    Institut f\"{u}r Theoretische Physik, 
        Universit\"{a}t Bern, \\
        Sidlerstrasse 5, CH-3012 Bern, Switzerland}
\vspace{1.8cm}

{\large\bf Abstract} 
\end{center} 
\vspace{5mm} 
\noindent 
We discuss recent calculations of the $O(\alpha)$ QED corrections to
the recoil electron energy spectrum in neutrino electron scattering,
and to the spectrum of the combined energy of the recoil electron and
a possible accompanying photon emitted in the scattering process. We
then examine the role of these corrections in the interpretation of
precise measurements from solar neutrino electron scattering
experiments.


\end{titlepage}

\section{Introduction}

The calculation of the neutrino electron scattering cross section has
a long history. The cross section for the process $\nu_{e} + e
\rightarrow \nu_{e} + e$ was first computed by Feynman and Gell-Mann
almost half a century ago within the framework of an effective
four--fermion V$-$A theory \cite{FG}. The QED corrections to this
cross section were calculated in 1964 by Lee and Sirlin \cite{LS}, and
shortly afterwards Ram \cite{Ram} extended their calculations by
including hard photon emission. A few years later, 't Hooft computed
the lowest order prediction to this differential cross section in the
Standard Model (SM) \cite{tH}. Since then, the radiative corrections
to this process, which plays a fundamental role in the study of
electroweak interactions, have been investigated by many authors,
focusing on various aspects of the problem \cite{ZKN,SU-GV-MS-AHKKM,
  AH,MSS,DB1,DB2,Bernabeu}.

't Hooft's early SM predictions were used by Bahcall to examine the
total cross section, energy spectrum and angular distribution of
recoil electrons resulting from the scattering with solar neutrinos
\cite{B87}. In a later work, Bahcall, Kamionkowski and Sirlin
performed a detailed investigation of the radiative corrections to
these recoil electron spectra and total cross sections \cite{BKS}.
Their results showed the importance of these corrections for the
analysis of precise solar $\nu$--$e$ scattering experiments,
particularly of those measuring the higher energy neutrinos that
originate from \B8 decay.  In a very recent article \cite{MP00}, we
extended previous calculations of the $O(\alpha)$ QED corrections to
the recoil electron energy spectrum and evaluated the corresponding
corrections to the differential cross section with respect to the
total combined energy of the recoil electron and a possible
accompanying photon. In the same paper, on which this note is based,
we examined the role of these two different radiative corrections in
the interpretation of precise measurements from solar neutrino
electron scattering experiments.

The energy spectrum of electrons from solar neutrino scattering, first
measured by the Kamiokande collaboration \cite{Kam}, provides
important information for the investigation of possible solutions to
the long standing solar neutrino problem, namely the large deficit in
the observed neutrino flux from the Sun with respect to the
theoretical predictions \cite{Ba2000}.  Indeed, while solar neutrino
fluxes are predicted by the standard solar model, the shapes of the
neutrino energy spectra were shown by Bahcall to be essentially
independent of all solar parameters \cite{Ba91}.  The shape of the
energy spectrum of recoil electrons resulting from the scattering with
solar neutrinos can be therefore accurately calculated in a solar
model-independent way and compared with the experimental measurements.
Distortions of the electron spectrum can be interpreted in terms of
neutrino oscillations. In particular, different flavor oscillation
solutions to the solar neutrino problem are reflected in
characteristic modifications of the shape of the electron spectrum and
precise theoretical predictions and experimental measurements are
needed to discriminate among them \cite{Ba97,Lisi}.  (See also
ref.~\cite{Andre} for a study of the possibility of determining the
flavor content of the low-energy solar neutrino flux based on the
analysis of the shapes of the recoil electron spectra.) We refer the
reader to ref.~\cite{Concha} for an updated global two- and
three-neutrino oscillation analysis of solar neutrino data which
includes the electron spectrum measured by the Super--Kamiokande
collaboration \cite{SK}.

In the following we examine  the $O(\alpha)$ QED corrections to the SM
prediction for neutrino electron scattering, with contributions
involving either neutral currents (as in the $\nu_{\mu,\tau} + e
\rightarrow \nu_{\mu,\tau} + e$ process) or a combination of neutral
and charged currents (as in the $\nu_{e} + e \rightarrow \nu_{e} + e$
process).  In this analysis we make the approximation of neglecting
terms of $O(q^2/\mw^2)$, where $q^2$ is the squared four-momentum
transfer and $\mw$ is the $W$ boson mass.  Within this approximation,
which is excellent for present experiments ($|q^2/\mw^2| \!\sim\! 1$
when the electron recoil energy $\sim 6 \times 10^3$ TeV!), the SM
radiative corrections to these processes can be naturally divided into
two classes. The first, which we will call ``QED'' corrections,
consist of the photonic radiative corrections that would occur if the
theory were a local four--fermion Fermi theory rather than a gauge
theory mediated by vector bosons; the second, which we will refer to
as the ``electroweak'' (EW) corrections, will be the remainder. This
split--up of the QED corrections is sensible as they form a finite
(both infrared and ultraviolet) and gauge--independent subset of
diagrams. We refer the reader to ref.~\cite{Si78-80} for a detailed
study of this separation.

The QED radiative corrections are due to both loop diagrams (virtual
corrections) and to the bremsstrahlung radiation (real photons)
accompanying the scattering process. Of course, only this combination
of virtual and real photon corrections is free from infrared
divergences. To order $\alpha$, the bremsstrahlung events correspond
to the inelastic process $\nu_l + e \rightarrow $ $\nu_l + e +\gamma $
($l=e,\mu$ or $\tau$). Experimentally, bremsstrahlung events in which
photons are too soft to be detected are counted as contributions to
the elastic scattering $\nu_l + e \rightarrow$ $\nu_l + e$. The cross
section for these events should be therefore added to the theoretical
prediction of the elastic cross section, thus removing its infrared
divergence.

We will divide the bremsstrahlung events into ``soft'' (hereafter SB)
and ``hard'' (hereafter HB), according to the energy of the photon
being respectively lower or higher than some specified threshold
$\epsilon$. We should warn the reader that the words ``soft'' and
``hard'' may be slightly deceiving. Indeed, if $\epsilon$ is large
(small), the SB (HB) cross section will also include events with
relatively high (low) energy photons.  While calculations of both soft
and hard bremsstrahlung are often performed under the assumption that
$\epsilon$ is a very small parameter, much smaller than the mass of
the electron or its final momentum, we will also discuss
results for the case in which $\epsilon$ is an arbitrary parameter
constrained only by the kinematics of the process.  Indeed, the HB
cross section (contrary to the SB one) is by itself, at least in
principle, a physically measurable quantity for any kinematically
allowed value of this threshold. All calculations have been carried
out without neglecting the electron mass.

In sect.~2 we present the lowest order prediction for the final
electron spectrum, together with its QED corrections (virtual, soft
and hard contributions). In sect.~3 we examine the spectrum of the
total combined energy of the recoil electron and a possible
accompanying photon emitted in the scattering process. We summarize
the main results in sect.~4.

\section{QED Corrections to the Electron Spectrum}

The SM prediction for the elastic neutrino electron differential
cross section is, in lowest--order and neglecting terms of
$O(q^2/\mw^2)$ \cite{tH},
\be
   \left[\frac{d\sigma}{dE}\right]_0 \;=\; \frac{2mG_{\mu}^2}{\pi}
        \left[\gl^2 +\gr^2 \left(1-z\right)^2 -\gl \gr 
        \left(\frac{m z}{\nu}\right)\right],
\label{eq:treelevel}
\ee
where $m$ is the electron mass, $G_{\mu}=1.16637(1) \times
10^{-5}\gev^{-2}$ is the Fermi coupling constant \cite{FOS}, $\gl =
\sin^2 \!\theta_{\smallw} \pm 1/2$ (upper sign for $\nu_e$, lower sign
for $\nu_{\mu,\tau}$), $\gr = \sin^2 \!\theta_{\smallw}$ and
$\sin\theta_{\smallw}$ is the sine of the weak mixing angle. In this
elastic process $E$, the electron recoil energy, ranges from $m$ to
$E_\smallmax =$ $[m^2 +(2\nu +m)^2]/[2(2\nu +m)]$, $z=(E-m)/\nu$ and
$\nu$ is the incident neutrino energy in the frame of reference in
which the electron is initially at rest.  We will refer to the $L$,
$R$ and $LR$ parts of an expression to indicate its terms proportional
to $\gl^2$, $\gr^2$ and $\gl \gr$, respectively. For example, the $R$
part of $\left[d\sigma/dE\right]_0$ (eq.~(\ref{eq:treelevel})) is
$(2mG_{\mu}^2/\pi) \gr^2 (1-z)^2$.

According to the definition discussed earlier, the one--loop QED
corrections to neutrino electron scattering consist of the photonic
vertex corrections (together with the diagrams involving the field
renormalization of the electrons) computed with the local
four--fermion Fermi Lagrangian.  These corrections give rise to the
following expression for the differential cross section:
\be 
   \left[\frac{d\sigma}{dE}\right]_{\rm Virtual}
   \;=\; \frac{2mG_{\mu}^2}{\pi} 
   \Biggl[\frac{\alpha}{\pi} \,\delta(E,\nu)\Biggr],
\label{eq:V}
\ee
where 
\bea \delta(E,\nu) &=& \gl^2\; \left\{V_1(E) +V_2(E)\left[
    z-1-\frac{mz}{2\nu} \right]
\right\} \nonumber  \\
&+& \gr^2\; \left\{V_1(E)\left(1-z\right)^2 + V_2(E)\left[
    z-1-\frac{mz}{2\nu} \right]
\right\} \nonumber  \\
&-& \gl\gr\,\left\{ \left[V_1(E)-V_2(E)\right] \left(\frac{mz}
    {\nu}\right) +2V_2(E) \left[z-1-z^2 \right] \right\}, 
\eea
\bea V_1(E) &=& \left(2\ln\!\frac{m}{\lambda}\right)
\left[1-\frac{E}{2l} \ln\!\left(\frac{E+l}{E-l}\right) \right] -2
-\frac{E}{l}\left[ \,\dilog\!\left(\frac{l-E+m}{2l}\right)
\right.     \nonumber  \\
&-& \left. \dilog\!\left(\frac{l+E-m}{2l}\right) \right] +
\frac{1}{4l} \left[3E+m-E\ln\!\left(\frac{2E+2m}{m}\right)\right]
\ln\!\left(\frac{E+l}{E-l}\right), \\ ~ \nonumber\\
V_2(E) &=& \frac{m}{4l}\ln\!\left(\frac{E+l}{E-l}\right).  
\eea
$\lambda$ is a small photon mass introduced to regularize the infrared
divergence and $l=\sqrt{E^2-m^2}$ is the three-momentum of the
electron. The dilogarithm $\dilog(x)$ is defined by
$$ \dilog(x) = -\int_0^x \!dt \,\frac{\ln(1-t)}{t}.  $$
The $L$ part of eq.~(\ref{eq:V}) (with $\gl=1$) is identical to the
formula for the one--loop photonic corrections to the $\nu_e +e
\rightarrow \nu_e +e$ differential cross section computed long ago in
the pioneering work of Lee and Sirlin \cite{LS} using the
effective four--fermion Fermi V$-$A Lagrangian. The analogous formula
for the reaction involving an anti--neutrino $\overline{\nu}_e$
(rather than a neutrino $\nu_e$) can be found in the same article and
coincides\footnote{with the exception of a minor typographical error
  in their eq.~22, where the square bracket multiplying 
  $I_{\rm rad}$ should be
  squared. We thank Alberto Sirlin for confirming this point.}  with
the $R$ part of eq.~(\ref{eq:V}) (with $\gr = 1$). This identity is
simply due to the fact that the cross section for antineutrinos in the
local V$-$A theory is the same as that for neutrinos calculated with a
V$+$A coupling. On the contrary, the $LR$ part of eq.~(\ref{eq:V}) has
clearly no analogue in the V$\pm$A theory, but can be derived very easily 
once the $L$ and $R$ parts are known.

The $\nu_e +e\rightarrow$ $\nu_e +e +\gamma$ differential cross
section with emission of a soft photon was computed in ref.~\cite{LS},
once again by using the effective four--fermion Fermi V$-$A
Lagrangian. It can be identified with the $L$ part (with $\gl=1$)
of the soft photon corrections to the tree level result in 
eq.~(\ref{eq:treelevel}). The $L$, $R$ and $LR$ parts of these corrections 
(with $\gl=\gr=1$) are however identical,
because the whole soft bremsstrahlung cross section is proportional to
its lowest--order elastic prediction.  We can therefore write the soft
photon emission cross section in the following factorized form:
\be
   \left[\frac{d\sigma}{dE}\right]_{\rm SB}
   \;=\; \frac{\alpha}{\pi} \;I_\gamma (E,\epsilon)
   \left[\frac{d\sigma}{dE}\right]_0,
\label{eq:SB}
\ee
with 
\bea
    I_\gamma(E,\epsilon) &=& \left(2\ln\!\frac{\lambda}{\epsilon}\right)
    \left[1-\frac{E}{2l} \ln\!\left(\frac{E+l}{E-l}\right) \right]
    + \frac{E}{2l}\left\{\,
    L\! \left(\frac{E+l}{E-l}\right) -L\! \left(\frac{E-l}{E+l}\right)
                  \right. \nonumber  \\
    &+& \left.    \ln\!\left(\frac{E+l}{E-l}\right) 
        \left[1-2\ln\!\left(\frac{l}{m}\right)\right] \right\} +1-2\ln\!2
\eea
and
$$
   L(x) = \int_0^x \!dt \,\frac{\ln|1-t|}{t}.  
$$ 
(For $x\in \mbox{I}\!\mbox{R}$, $L(x) = -\mbox{Re} [\dilog(x)]$.)
This result is valid under the assumption that $\epsilon$, the maximum
soft photon energy, is much smaller than $m$ or the final momentum of
the electron.  As we mentioned earlier, in the following we will
discuss numerical results for the case in which $\epsilon$ is an
arbitrary parameter.  The reader will notice that the sum
$(V_1(E)+I_\gamma(E,\epsilon))$ does not depend on $\lambda$, the
infrared regulator. Indeed, the infrared divergence of the virtual
corrections (eq.~(\ref{eq:V})) is canceled by that arising from the
soft photon emission (eq.~(\ref{eq:SB})).

The SM prediction for the differential neutrino electron cross
section 
\be
     \nu_l + e \rightarrow \nu_l + e \;(+\gamma), 
\label{eq:nuegamma}
\ee
where $(+\gamma)$ indicates the possible emission of a photon, can be
cast, up to corrections of $O(\alpha)$, in the following form:
\bea
   \left[\frac{d\sigma}{dE}\right]_{\rm SM}
   & = & \frac{2mG_{\mu}^2}{\pi} 
        \Biggl\{\gl^2(E) \left[1+\frac{\alpha}{\pi} \fl(E,\nu) \right]
        +\gr^2(E) \left(1-z\right)^2 
        \left[1+\frac{\alpha}{\pi} \fr(E,\nu) \right] \nonumber\\
   & &  -\gl(E) \gr(E) \left(\frac{m z}{\nu}\right)
        \left[1+\frac{\alpha}{\pi} \flr(E,\nu) \right] \Biggr\}.
\label{eq:SMdE}
\eea 
(We remind the reader that terms of $O(q^2/\mw^2)$ are neglected
in our analysis.)  The deviations of the functions $\gl(E)$ and
$\gr(E)$ from the lowest--order values $\gl$ and $\gr$ reflect the
effect of the electroweak corrections, which have been studied by
several authors \cite{SU-GV-MS-AHKKM,MSS,BKS}. (See ref.~\cite{BKS}
for simple numerical results.)

The functions $\fx(E,\nu)$ ($X=L,R$ or $LR$) describe the QED effects
(real and virtual photons). For simplicity of notation their $\nu$
dependence will be dropped in the following. Each of these functions
is the sum of virtual (V), soft (SB) and hard (HB) corrections,
\be
    \fx(E) = \fx^{\sss V}(E) + 
             \fx^{\sss SB}(E,\epsilon) + 
             \fx^{\sss HB}(E,\epsilon).
\label{eq:fxsum}
\ee
The analytic expressions for $\fx^{\sss V}(E)$ and $\fx^{\sss
  SB}(E,\epsilon)$ can be immediately read from eqs.~(\ref{eq:V}) and
(\ref{eq:SB}) respectively (the latter being valid only in the small
$\epsilon$ limit) and their sums, which are infrared--finite, will be
denoted by
\be
    \fx^{\sss VS}(E,\epsilon) = \fx^{\sss V}(E) +
                          \fx^{\sss SB}(E,\epsilon).
\ee
Analytic expressions from which one can obtain $\fl^{\sss
  HB}(E,\epsilon)$ and $\fr^{\sss HB}(E,\epsilon)$ were calculated
long ago by Ram in the small $\epsilon$ approximation, keeping the
logarithmically divergent terms proportional to $\ln(\epsilon/m)$ but
neglecting the remaining $\epsilon$--dependent terms \cite{Ram}.  The
formulae are nonetheless long and complicated. The function
$\flr^{\sss HB}(E,\epsilon)$ was computed only very recently
\cite{MP00}.

In ref.~\cite{MP00} we created {\tt BC}, a combined {\tt
  Mathematica}--{\tt FORTRAN} code\footnote{The code {\tt BC},
  available upon request, computes all QED corrections discussed in
  this note.} to compute the $\fx^{\sss HB}(E,\epsilon)$ functions for
arbitrary positive values of the parameter $\epsilon$ up to the
kinematic limit $\nu$ ($\nu$, the incident neutrino energy in the
laboratory system, is also the maximum possible energy of the emitted
photon). We refer the interested reader to this paper for details
concerning this computation and the comparison with Ram's results.
Our results confirm Ram's ones in the small $\epsilon$ limit. If
$\epsilon$ is not small, the discrepancy between them can become very
large. Moreover, Ram's results for $\fl^{\sss HB}(E,\epsilon)$ and
$\fr^{\sss HB}(E,\epsilon)$ are not always positive. This is of course
an unphysical property because the HB differential cross section,
being a transition probability for a physical process, cannot be
negative. Our functions $\fl^{\sss HB}(E,\epsilon)$ and $\fr^{\sss
  HB}(E,\epsilon)$ are always positive (or zero).

The total $O(\alpha)$ QED corrections $\fl(E)$ and $\fr(E)$, given by
the sum of V, SB and HB contributions (see eq.~(\ref{eq:fxsum})), can
be easily obtained by adding the analytic results of eqs.~(\ref{eq:V})
and (\ref{eq:SB}) to Ram's HB (lengthy) ones. Both SB and HB
corrections were computed in the small $\epsilon$ approximation, and
the logarithmically divergent terms proportional to $\ln(\epsilon/m)$
exactly drop out upon adding these soft and hard contributions. The
remaining $\epsilon$--dependent terms, which were neglected in both SB
and HB calculations, must cancel in the sum as well, and are therefore
irrelevant in the computation of the total QED corrections of
eq.~(\ref{eq:SMdE}). The $LR$ case is slightly different: Ram's
formulae, which were used to derive the small $\epsilon$ approximation
for $\fl^{\sss HB}(E,\epsilon)$ and $\fr^{\sss HB}(E,\epsilon)$, do
not provide us with the corresponding $LR$ correction. In order to
compute $\flr(E)$ we have therefore added the V and SB analytic
results of eqs.~(\ref{eq:V}) and (\ref{eq:SB}) to our HB numerical
results.  The ``exact'' $\epsilon$ dependence of our HB results is not
completely canceled by that of the SB, which includes only terms
proportional to $\ln(\epsilon/m)$, and the sum $\flr(E)$ contains
therefore a residual (not logarithmically divergent) dependence on the
photon energy threshold $\epsilon$. This spurious dependence has been
minimized by fixing $\epsilon$ to be a very small value
$\epsilon_{\smalll \smallr}$ chosen so as to have an estimated induced
relative error as small as $O(0.1\%)$ \footnote{with the exception of
  $E$ belonging to a tiny interval of $O(\epsilon_{\smalll \smallr})$
  at the endpoint $E_\smallmax$. Note that $0.1\%$ is also the relative
  numerical uncertainty used by our code {\tt BC} in the computation
  of the functions $\fx^{\sss HB}(E,\epsilon)$ and produces a totally
  negligible relative error $(\alpha/\pi)\fx^{\sss
    HB}(E,\epsilon)\times 0.1\%$ in the corresponding $X$ parts of the
  differential cross section in eq.~(\ref{eq:SMdE}).}.

Eqs.~(\ref{eq:V}) and (\ref{eq:SB}) determine the analytic expression
of $\fx^{\sss VS}(E,\epsilon)$ (the infrared--finite sum of V and SB
corrections) in the small $\epsilon$ approximation. But the complete
$\epsilon$ dependence of our numerical $\fx^{\sss HB}(E,\epsilon)$
computations, combined with the knowledge of the above described
$\fx(E)$ functions, allows us to determine also the ``exact''
$\fx^{\sss VS}(E,\epsilon)$ functions via the subtraction
\be
\fx^{\sss VS}(E,\epsilon) =\fx(E)-\fx^{\sss HB}(E,\epsilon).
\label{eq:vsexact}
\ee
These will be the ``exact'' VS corrections employed in the rest of our
analysis.

In fig.~1 we plotted the functions $\fx(E)$ (thick solid), $\fx^{\sss
  HB}(E,\epsilon)$ (medium solid) and $\fx^{\sss VS}(E,\epsilon)$
(thin solid) for $\nu=0.862 \mev$. The threshold $\epsilon$ in the VS
and HB functions was set to $0.02\mev$. In fig.~2 we plotted the same
functions with $\nu=10 \mev$ and $\epsilon=1\mev$.  These two values
of the neutrino energy were chosen for their relevance in the study of
solar neutrinos: $\nu=0.862\mev$ is the energy of the (almost)
monochromatic neutrinos produced by electron capture on \Be7 in the
solar interior, while $\nu = 10\mev$ belongs to the continuous energy
spectrum of the solar neutrinos that originate from the decay of
\B8 $\!\!$.            
In figs.~1 and 2 we also plotted the simple approximate formulae for
$\fx(E)$ introduced in ref.~\cite{BKS} (dotted lines). These compact
analytic expressions were obtained by modifying the expressions of
ref.~\cite{MSS}, which had been evaluated in the extreme relativistic
approximation. (The $LR$ term of the differential cross section, being
proportional to $(m/\nu)$, vanishes in the extreme relativistic limit
and, therefore, cannot be derived from ref.~\cite{MSS}. As a
consequence, the $LR$ approximation of ref.~\cite{BKS} is only a (very
educated!)  guess.)  Thanks to their simplicity, the compact formulae
of ref.~\cite{BKS} are easy to use and are employed, for example, by
the Super--Kamiokande collaboration in their Monte Carlo simulations
for the analysis of the solar neutrino energy spectrum.

As it was noted in refs.~\cite{Ram, BKS}, all $\fx(E)$ functions contain a
term which diverges logarithmically at the end of the spectrum. This
feature, related to the infrared divergence, is similar to the one
encountered in the QED corrections to the $\mu$--decay spectrum
\cite{BFS,KS}. If $E$ gets very close to the endpoint we have
$(\alpha/\pi)\fx(E) \approx -1$, clearly indicating a breakdown of the
perturbative expansion and the need to consider multiple-photon
emission. However, this divergence
can be easily removed (in agreement with the KLN theorem
\cite{KS,KLN}) by integrating the differential cross section over
small energy intervals corresponding to the experimental energy
resolution. We also note that the singularity of
$\flr(E)$ for $E=m$ does not pose a problem, as the product
$(mz/\nu)\flr(E)$, which appears in the $LR$ part of the differential
cross section, is finite in the limit $E\rightarrow m$.  This can be
seen from the dashed line in the $LR$ plot of fig.~1, which indicates 
the function $(mz/\nu)\flr(E)$. In the same plot, the dot-dashed line
is the product of the $\flr(E)$ approximation of ref.~\cite{BKS} and 
$(mz/\nu)$.

\section{Spectrum of the Combined Energy of Electron and Photon}

We will now turn our attention to the analysis of the differential
$\nu_l + e \rightarrow$ $\nu_l + e \;(+\gamma)$ cross section relevant
to experiments measuring the {\em total combined energy} of the recoil
electron and a possible accompanying photon emitted in the scattering
process. We will begin by considering bremsstrahlung events with a
photon of energy $\omega$ larger than the usual threshold $\epsilon$
(HB).

The HB differential cross section with respect to the sum of the
electron and photon energies is computed by our code {\tt BC} (see
ref.~\cite{MP00}). The code first calculates the HB corrections to the
energy spectrum of the final neutrino; the HB differential cross
section $[d\sigma/d(E+\omega)]_{\rm HB}$ is then immediately derived
via energy conservation.  A check of the consistency of our results
was performed by comparing the values of the total HB cross section
$\sigma_{\rm HB}(\nu,\epsilon)$ obtained by integrating both
$[d\sigma/d(E+\omega)]_{\rm HB}$ and the differential HB cross section
of sect.~2 for several values of $\nu$ and $\epsilon$.  All relative
deviations were found to be smaller than 0.1\% (which is also the
relative accuracy of the integrands).  In the elastic reaction $\nu_l
+ e \rightarrow$ $\nu_l + e$, the final neutrino energy $\nu'$ ranges
from $\nu'_\smallmin=$ $\nu m/(2\nu+m)$ to $\nu'_\smallmax=\nu$ (the
value $\nu'=$ $\nu'_\smallmin$ occurs when the final electron and
neutrino are scattered back to back, with the electron moving in the
forward direction with $E=E\smallmax$; the value $\nu'=$
$\nu'_\smallmax$ occurs in the forward scattering situation).  When a
photon of energy $\omega > \epsilon$ is emitted, $\nu'$ varies between
$0$ and $\nu-\epsilon$, while the variable $E+\omega$ varies between
$m+\epsilon$ and $m+\nu$ (note that $m+\nu = E_\smallmax +
\nu'_\smallmin$).

How do we combine virtual, soft and hard bremsstrahlung contributions
in order to evaluate the complete $O(\alpha)$ QED prediction for the
differential cross section $d\sigma/d(E+\omega)$ of reaction
(\ref{eq:nuegamma})? In sect.~2 we computed the total QED corrections
by simply adding these three parts. Their sum does not depend on the
threshold $\epsilon$. The combination of VS and HB terms requires here
a more careful treatment for which we refer the interested reader,
once again, to ref.~\cite{MP00}. We will only discuss the results of
this analysis, which can be summarized in the following simple way.
Let's consider an experimental setup for $\nu$--$e$ scattering able to
measure the photon energy if it's higher than a threshold $\epsilon$,
but completely blind to low energy photons $(\omega\!  <\!\epsilon)$.
Let's also assume that the electron energy $E$ is precisely measurable
independently of its value.  This detector can measure the usual
electron spectrum $d\sigma/dE$ as well as the differential cross
section $d\sigma/dE_\omega$, where the variable $\eom$ is defined as
follows,
\be 
    E_\omega \equiv \left\{
    \begin{array}{ll}
    E+\omega & \quad\mbox{if} \;\; \omega \geq \epsilon \\
    E & \quad\mbox{if} \;\; \omega < \epsilon
                \end{array} \right..
\label{eq:eom}
\ee
The SM prediction for the spectrum of the combined energy of electron
and photon in reaction (\ref{eq:nuegamma}) can be cast, up to
corrections of $O(\alpha)$, in the form
\bea
        \left[\frac{d\sigma}{dE_\omega}\right]_{\rm SM} \!\!\!\!
      &=& \!\!\frac{2mG_{\mu}^2}{\pi}
        \Biggl\{\gl^2(\eom) \!\left[\theta 
          +\frac{\alpha}{\pi} \flov(\eom,\epsilon,\nu) \right]
        +\gr^2(\eom) \left(1-\zom\right)^2 
        \!\left[\theta
        +\frac{\alpha}{\pi}\frov(\eom,\epsilon,\nu)\right]
                                \nonumber\\
      & & -\gl(\eom)\gr(\eom) \left(\frac{m \zom}{\nu}\right)
        \!\left[\theta+ \frac{\alpha}{\pi}
        \flrov(\eom,\epsilon,\nu)\right]\Biggr\},
\label{eq:SMdeom}
\eea
where $\zom=(\eom-m)/\nu$ and $\theta = \theta(E_\smallmax -
E_\omega)$.  As we mentioned in sect.~2, the deviations of the
functions $\gl(\eom)$ and $\gr(\eom)$ from the lowest--order values
$\gl$ and $\gr$ reflect the effect of the electroweak corrections (for
virtual corrections it is $\omega=0$ and $\eom=E$). The functions
$\fxov(\eom,\epsilon,\nu)$ ($X=L,R$ or $LR$), defined in the range
$[m, m+\nu]$, describe the QED effects and can be written in the very
simple form (once again, for simplicity of notation, we will drop
their $\nu$ dependence)
\be
     \fxov(\eom,\epsilon) = \fx^{\sss VS}(\eom,\epsilon) + 
                            \fxov^{\sss HB}(\eom,\epsilon),
\label{eq:fxovsum}
\ee
where $\fx^{\sss VS}(\eom,\epsilon)$ are the ``exact'' VS corrections
of sect.~2 (eq.~(\ref{eq:vsexact})) and the functions $\fxov^{\sss
  HB}(\eom,\epsilon)$ are derived by dividing the $L$, $R$ and $LR$
parts of the above--mentioned HB cross section
$[d\sigma/d(E+\omega)]_{\rm HB}$ by $C\gl^2$, $C\gr^2(1-\zom)^2$ and
$-C\gl\gr(m \zom/\nu)$ respectively, with $C=2mG_{\mu}^2\alpha/\pi^2$.
The $\theta$ functions in eq.~(\ref{eq:SMdeom}) reflect the fact that
the lowest order prediction for $d\sigma/dE_\omega$ has a step at
$\eom = E_\smallmax$ and is zero if $\eom$ lies outside the elastic
range $[m, E_\smallmax]$.  The VS functions $\fx^{\sss
  VS}(\eom,\epsilon)$ are proportional to the same $\theta$ function,
while the corrections $\fxov^{\sss HB}(\eom,\epsilon)$ are set to zero
if $\eom \notin\; [m+\epsilon, m+\nu]$.  We would like to emphasize
the $\epsilon$ dependence of the complete QED corrections
$\fxov(\eom,\epsilon)$, to be contrasted with the $\epsilon$
independence of the $\fx(E)$ functions of sect.~2.

In figs.~3 and 4 we compare the results of sects.~2 and 3. In
fig.~3 we chose $\nu=0.862\mev$ and plotted the functions $\fx(E)$
(thick) and $\fxov(\eom,\epsilon)$ for $\epsilon=$ 0.1 MeV (medium)
and 0.001 MeV (thin). In fig.~4 we plotted the same functions
with $\nu=10\mev$ ($\epsilon=$ 1 MeV, 0.1 MeV).

We would like to remind the reader that the functions $\fx(E)$ can be
obtained from $\fxov(\eom,\epsilon)$ by simply setting $\epsilon=\nu$.
The limiting case $\epsilon=0$ was studied in detail in
ref.~\cite{DB2} (in particular, the results of the second article of
this reference were obtained, like ours, without employing the
ultrarelativistic approximation $E \gg m$).

\section{Discussion and Conclusions}

When are the results of sects.~2 and 3 applicable? In sect.~2 we
presented the $O(\alpha)$ SM prediction for the electron spectrum in
the reaction $\nu_l + e \rightarrow \nu_l + e \;(+\gamma)$
(eq.~(\ref{eq:SMdE})), where $(+\gamma)$ indicates the possible
emission of a photon. In this calculation we assumed that the
final--state photon is not detected and, as a consequence, we
integrated over all possible values of the photon energy $\omega$.
Therefore, eq.~(\ref{eq:SMdE}) is the appropriate theoretical
prediction to use in the analysis of $\nu$--$e$ scattering when the
detector is completely blind to photons of all energies, but can
precisely measure $E$, the energy of the electron. Of course, a
detector could provide more information by detecting photons as soon
as their energy is above an experimental threshold $\epsilon$.  In
this case, still assuming a precise determination of $E$, one can
employ eq.~(\ref{eq:SMdE}), minus its HB correction, to analyze those
events which are counted as nonradiative (elastic), while the HB part
can be used, at least in principle, for a separate determination of
the inelastic cross section. Indeed, contrary to previous
calculations, our predictions are valid for an arbitrary value of the
threshold $\epsilon$ (and include the previously unknown $LR$ term).

In sect.~3 we examined the spectrum of the total combined energy of
the recoil electron and a possible accompanying photon emitted in the
scattering process (eq.~(\ref{eq:SMdeom})). This type of analysis is
useful when the photon energy $\omega$ cannot be separately determined
although it fully contributes to the precise total energy measurement
if its value is above a specific threshold $\epsilon$. Let's consider
an experimental setup able to measure the photon energy if it's higher
than $\epsilon$, but completely blind to low energy photons $(\omega\!
<\!\epsilon)$. Let's also assume that the electron energy $E$ is
precisely measurable independently of its value. This detector can
determine both the differential cross section $d\sigma/dE_\omega$
(eq.~(\ref{eq:SMdeom})) and the electron spectrum $d\sigma/dE$
(eq.~(\ref{eq:SMdE})) (as well as its separate HB component). There
are experiments, however, which cannot measure $E$, but only $\eom$,
with a specific value of the threshold $\epsilon$. BOREXINO \cite{BX}
and KamLAND \cite{KL}, for example, are liquid scintillation detectors
in which photons and electrons induce practically the same response.
If a photon is emitted in the $\nu$--$e$ scattering process, its
energy $\omega$ is counted together with $E$, provided their sum lies
within a specific range. The appropriate theoretical prediction for
their analysis is given, therefore, by the cross section
$d\sigma/dE_\omega$ of eq.~(\ref{eq:SMdeom}) with a very small value
of $\epsilon$ (for the case $\nu=0.862\mev$ see the thin lines in
fig.~3). However, we should point out that although the QED
corrections $\fx(E)$ (in eq.~(\ref{eq:SMdE})) and
$\fxov(\eom,\epsilon)$ with small $\epsilon$ (in
eq.~(\ref{eq:SMdeom})) are different, their numerical values are very
small when $\nu=0.862\mev$, the energy of the monochromatic neutrinos
produced by electron capture on \Be7 in the solar interior.  In fact,
as shown in fig.~3, both $(\alpha/\pi)\fx(E)$ and
$(\alpha/\pi)\fxov(\eom,\epsilon)$ with small $\epsilon$ are in this
case of $O(\lsim 1\%)$, and neither of the above collaborations is
likely to reach this high level of accuracy in their analyses of the
crucial \Be7 line.

There are detectors in which it might not be possible to identify the
measured energy with either $E$ or $\eom$. Indeed, the electron and
the photon may produce indistinguishable signals and the total
observed energy might not be the simple sum of $E$ and $\omega$, but
some other function of these two variables. Super--Kamiokande (SK),
for example, a water Cherenkov counter measuring the light emitted by
electrons recoiling from neutrino scattering, uses the number of hit
photomultiplier tubes to determine the electron energy. However, a
photon emitted in the scattering process may induce additional hits
indistinguishable from those of the electron. Moreover, a photon and
an electron of the same energy may produce different numbers of hits
and, therefore, it might not be possible to identify the total
measured energy with the sum $E+\omega$.

SK measures solar neutrinos with energies varying from 5 to 18 MeV.
For $\nu=10\mev$, fig.~4 shows that the QED corrections to the
differential cross sections $d\sigma/dE$ (eq.~(\ref{eq:SMdE})) and
$d\sigma/dE_\omega$ (eq.~(\ref{eq:SMdeom})) are of $O(1\%)$.
Corrections of this order may be relevant for the analysis of the very
precise data obtained by this collaboration. In fact, SK's Monte Carlo
simulations of the expected energy spectrum of recoil electrons from
solar neutrino scattering include the QED corrections of
ref.~\cite{BKS} (as well as the EW ones). As we investigated 
in sect.~2, these corrections provide good approximations of the
complete $O(\alpha)$ QED corrections $\fx(E)$ to the electron spectrum
of eq.~(\ref{eq:SMdE}) (see fig.~2). Our previous discussion, however,
seems to suggest that these corrections are not appropriate for SK's
solar neutrino analysis.  On the other hand, the SM prediction for the
spectrum of the combined energy of electron and photon of sect.~3
(eq.~(\ref{eq:SMdeom})) may be suitable, but only if we can assume a
similar efficiency in the detection of photons and electrons, and if
also relatively low energy electrons contribute to the total energy
measurement.  If these conditions are not met, and the precision of
the data requires it, one should probably perform a dedicated analysis
of the double differential cross section $d^2\sigma/(dE \,d\omega)$
with a response function specifically designed for this detector. A
triple differential cross section $d^3\sigma/(dE \,d\omega \,d\phi)$,
where $\phi$ is the angle between the directions of the electron and
the photon, may also be useful (see the first article of
ref.~\cite{Bernabeu}).

\vspace{0.3cm}

\noindent I would like to thank Massimo Porrati for organizing this
very pleasant and interesting symposium, and Alberto Sirlin for
innumerable instructive discussions on the topic presented here.


\newpage

\newpage
\begin{figure}[tbp]
\vspace{-2.8cm}\hspace{-3cm}\includegraphics[width=20cm]{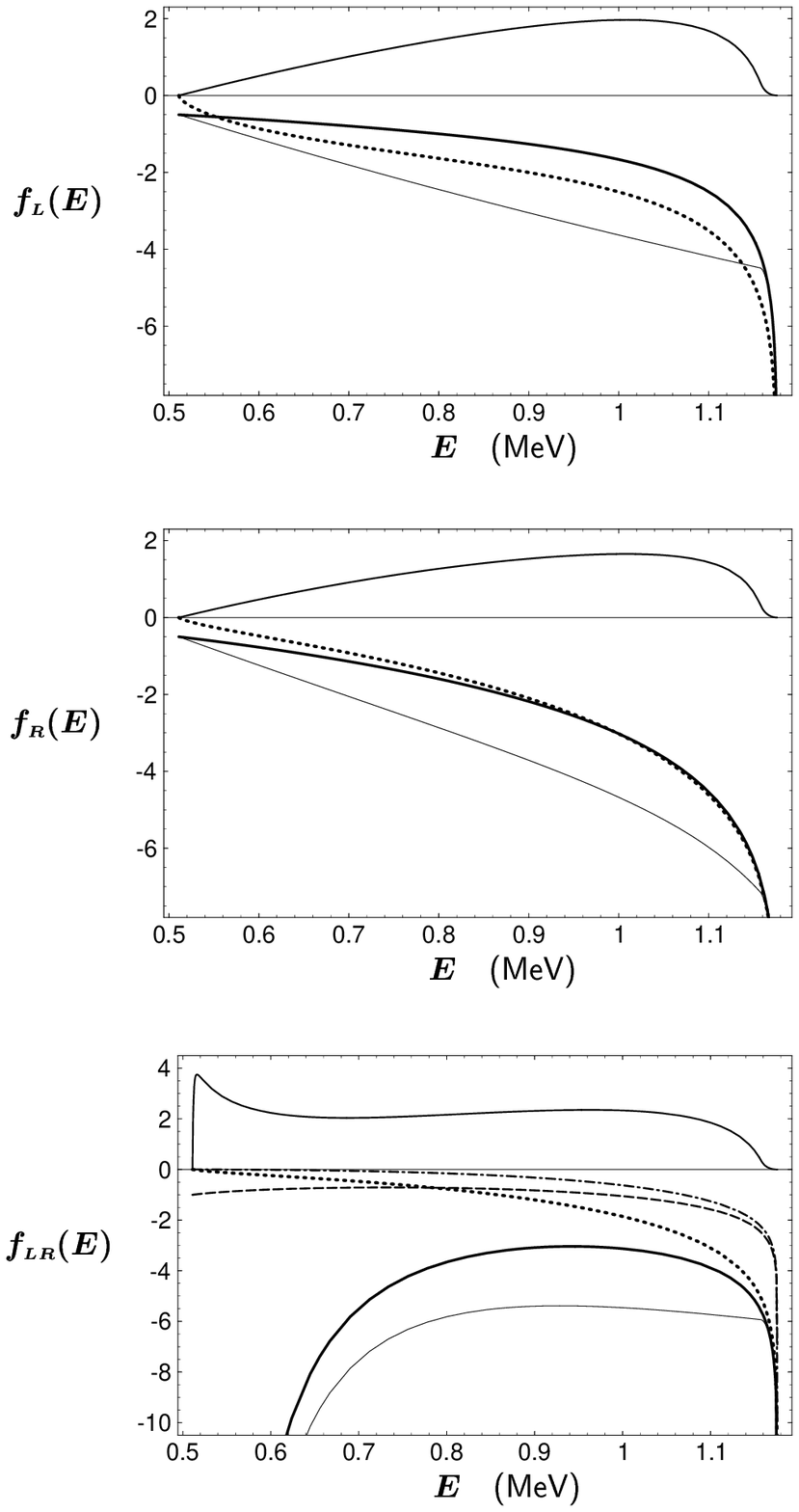}
\vspace{-7.5cm}\caption{\sf 
  The functions \bm{$\fx(E)$} (thick solid), \bm{$\fx^{\sss
      HB}(E,\epsilon)$} (medium solid) and \bm{$\fx^{\sss
      VS}(E,\epsilon)$} (thin solid) for \bm{$\nu=$} 0.862 MeV and
  \bm{$\epsilon=$} 0.02 MeV. The dotted lines represent the
  \bm{$\fx(E)$} approximations of ref.~\cite{BKS}. In the $LR$ figure, 
  the dot-dashed line is the product of the \bm{$\flr(E)$}
  approximation of ref.~\cite{BKS} and \bm{$(mz/\nu)$},
  while the dashed line indicates the product \bm{$(mz/\nu)\flr(E)$}.}
\label{figure:f1}
\end{figure}

\newpage
\begin{figure}[tbp]
\vspace{-2.8cm}\hspace{-3cm}\includegraphics[width=20cm]{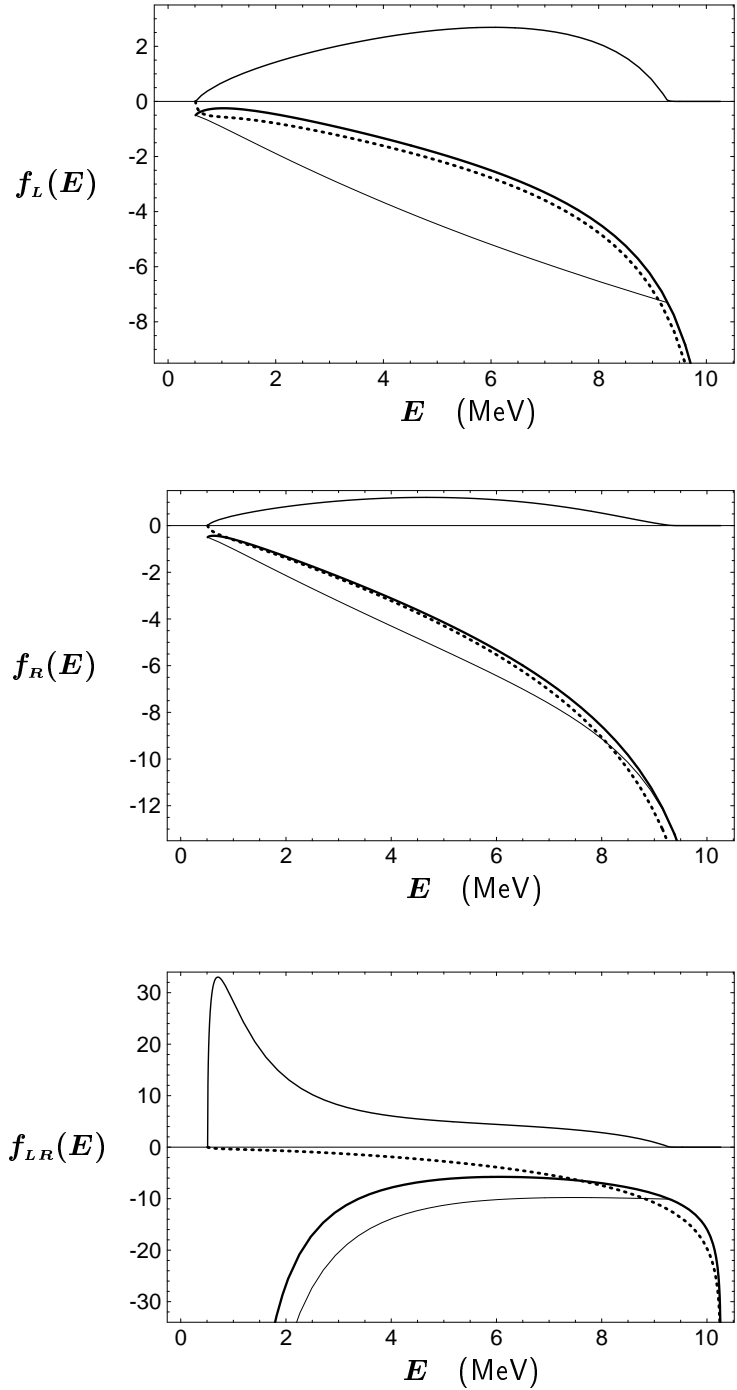}
\vspace{-7.5cm}\caption{\sf 
  Same as Fig.~\ref{figure:f1}, but for \bm{$\nu=$} 10 MeV and
  \bm{$\epsilon=$} 1 MeV.  The dashed and dot-dashed lines are very
  close to zero and are not indicated.}
\label{figure:f2} 
\end{figure}

\newpage
\begin{figure}[tbp]
\vspace{-2.8cm}\hspace{-2.5cm}\includegraphics[width=20cm]{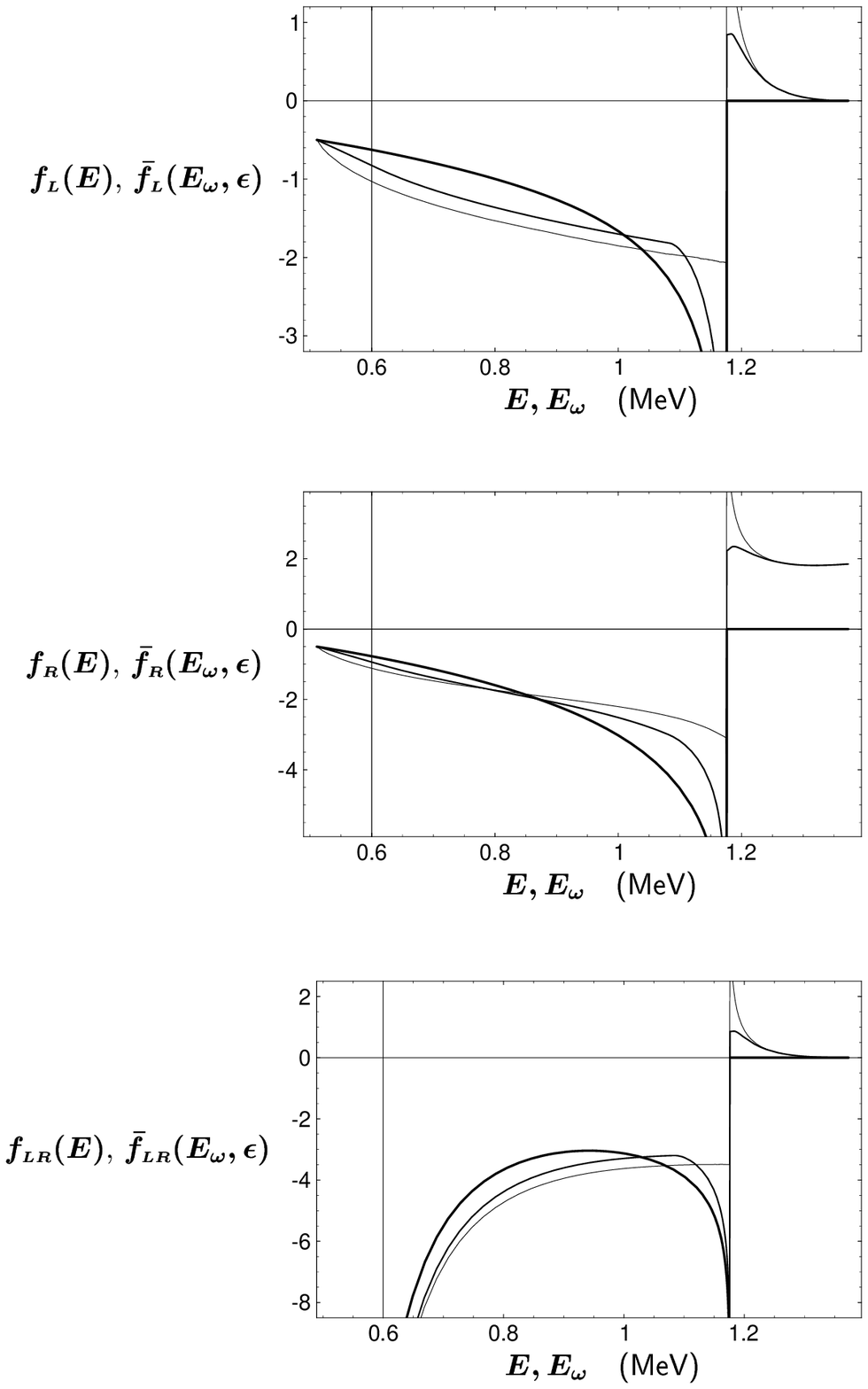}
\vspace{-7.5cm}\caption{\sf The functions \bm{$\fx(E)$} (thick) and 
  \bm{$\fxov(\eom,\epsilon)$} for \bm{$\epsilon=$} 0.1 MeV (medium)
  and 0.001 MeV (thin). \bm{$\nu=$} 0.862 MeV.}
\label{figure:f3}
\end{figure}

\newpage
\begin{figure}[tbp]
\vspace{-2.8cm}\hspace{-2.5cm}\includegraphics[width=20cm]{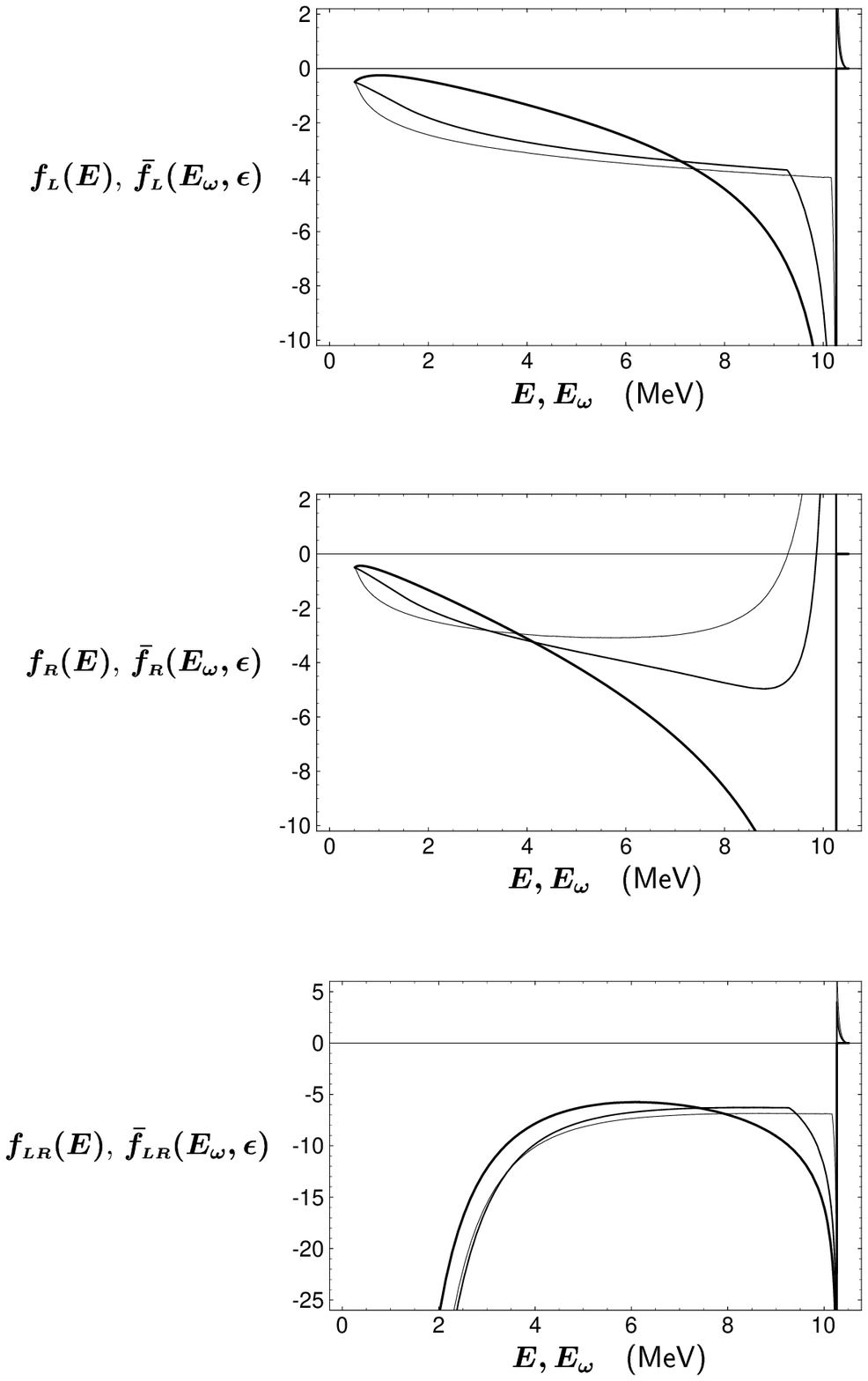}
\vspace{-7.5cm}\caption{\sf \sf The functions \bm{$\fx(E)$} (thick) and 
  \bm{$\fxov(\eom,\epsilon)$} for \bm{$\epsilon=$} 1 MeV (medium) and
  0.1 MeV (thin). \bm{$\nu=$} 10 MeV.}
\label{figure:f4}
\end{figure}

\end{document}